\documentclass[aps,twocolumn,showpacs]{revtex4}
\usepackage[psamsfonts]{amsfonts,amssymb}
\usepackage{bm}
\usepackage{graphicx}

\begin{document}

\title{Reducing or enhancing chaos using periodic orbits}
\author{R. Bachelard$^1$, C. Chandre$^1$, X. Leoncini$^{1,2}$ }
\affiliation{$^1$
Centre de Physique Th\'eorique\footnote{Unit\'e Mixte de Recherche (UMR 6207) du CNRS, et des universit\'es Aix-Marseille I, Aix-Marseille II et du Sud Toulon-Var. Laboratoire affili\'e \`a la
FRUMAM (FR 2291)}, CNRS Luminy, Case 907, F-13288 Marseille cedex 09, France\\
$^2$ PIIM, Universit\'e de Provence-CNRS, Centre Universitaire de Saint-J\'er\^ome, F-13397 Marseille, France}
\pacs{05.45.-a}

\date{\today}
\begin{abstract}
A method to reduce or enhance chaos in Hamiltonian flows with two degrees of freedom is discussed. This method is based on finding a suitable perturbation of the system such that the stability of a set of periodic orbits changes (local bifurcations). Depending on the values of the residues, reflecting their linear stability properties, a set of invariant tori is destroyed or created in the neighborhood of the chosen periodic orbits. An application on a paradigmatic system, a forced pendulum, illustrates the method.
\end{abstract}
\maketitle

\textbf{
Changing the dynamical properties of a system is central to the design and performance of advanced devices based on many interacting particles. For instance, in particle accelerators, the aim is to find the appropriate magnetic elements to obtain an optimal aperture in order to increase the luminosity of the beam, thus requiring the decrease of the size of chaotic regions. In plasma physics, the situation is slightly more complex~: Inside a fusion device (like a tokamak or a stellarator), one needs magnetic surfaces in order to increase confinement. These surfaces are invariant tori of some fictitious time dynamics. A control strategy would be to recreate such magnetic surfaces by an appropriate modification of the apparatus (magnetic perturbation caused by a set of external coils). On the opposite, in order to collect energy and to protect the wall components, an external modification of the magnetic equilibrium has to be performed such that there is a highly chaotic layer at the border (like an ergodic divertor). Therefore these devices require a specific monitoring of the volume of bounded magnetic field lines. Another example is afforded by chaotic advection in hydrodynamics~: In the long run to achieve high mixing in microfluidics and microchannel devices in particular, the presence of regular region prevents such mixing, and hence a possible way to enhance mixing is to perturb externally the system according to some theoretical prescriptions, in order to destroy invariant surfaces.}

\section{Introduction}
\label{sec1}

A very fruitful information on the dynamics can be gained from the study of periodic orbits~\cite{Bcvit05}. First, because these particular orbits are generically almost everywhere in phase space, and second because they can be computed easily, i.e.\ with some short integration time. These periodic orbits together with their stability organize locally the dynamics. It is then natural to consider them as a cornerstone of control strategies. For instance, in order to create invariant tori of Hamiltonian systems, Cary and Hanson~\cite{hans84,cary86} proposed a method based on the computation of an indicator of the linear stability of a set of periodic orbits, namely Greene's residue~\cite{gree79}. It provides an algorithm to find the appropriate values of some pre-defined parameters in order to reconstruct invariant tori by vanishing some selected residues. First developed for two-dimensional symplectic maps, it has been extended to four dimensional symplectic maps, and has been applied to stellarators~\cite{hans94} (where periodic orbits are closed magnetic field lines) and particle accelerators~\cite{wan98}.  
 
In this article, we review and extend this residue method. The aim is to tune appropriately the parameters of the system such that appropriate bifurcations occur. It is well-known in the literature that local bifurcations occur when the tangent map associated with the Poincar\'e map obtained by a transversal intersection of the flow, has an eigenvalue which is a root of the unity. In particular, periodic orbits can lose their stability in case of multiple eigenvalues on the unit circle, i.e.\ when these eigenvalues are equal to 1 or $-1$ for two-dimensional symplectic maps. Therefore it is natural to consider Greene's residues as a way to locate those bifurcations. In this context, vanishing residues indicate the specific values of the parameters where significant change occurs in the system and hence will be the basis for the reduction of chaos (by creation of invariant tori) as in Refs.~\cite{hans84,cary86} but also for the destruction of regular structures.

In Sec.~\ref{sec2}, we review some basic notions on periodic orbits of Hamiltonian systems and their stability, and we explain the details of the residue method. We give the condition on the residues of a pair of Birkhoff periodic orbits to create an invariant torus in their vicinity, and a similar condition which leads to a destruction of nearby invariant tori. In Sec.~\ref{sec3}, we apply this method to the destruction and creation of librational and rotational invariant tori of a particular Hamiltonian system, a forced pendulum with two interacting primary resonances, used as a paradigm for the transition to Hamiltonian chaos.  

\section{The residue method}
\label{sec2}
We consider an autonomous Hamiltonian flow with two degrees of freedom which depends on a set of parameters denoted ${\bm \alpha}=(\alpha_1,\alpha_2,\ldots,\alpha_m)\in{\mathbb R}^m$~:
$$
\dot{z}={\mathbb J}\nabla H(z;{\bm\alpha}),
$$
where $z=({\mathbf p},{\mathbf q})\in{\mathbb R}^4$ and ${\mathbb J}=\left( \begin{array}{cc} 0 & -{\mathbb I}_2 \\ {\mathbb I}_2 & 0 \end{array}\right)$, and ${\mathbb I}_2$ being the two-dimensional identity matrix. In order to determine the periodic orbits of this flow and their linear stability properties, we also consider the tangent flow written as
$$
\frac{d}{dt}{J^t}(z)={\mathbb J}\nabla^2 H(z;{\bm\alpha}) J^t,
$$ 
where $J^0={\mathbb I}_4$ and $\nabla^2 H$ is the Hessian matrix (composed by second derivatives of $H$ with respect to its canonical variables). For a given periodic orbit with period $T$, the spectrum of the monodromy matrix $J^T$ gives its linear stability property. As the flow is volume preserving, the determinant of such a matrix is equal to 1. Moreover, if $\Lambda$ is an eigenvalue, so are $1/\Lambda$, $\Lambda^*$ and $1/\Lambda^*$. As the orbit is periodic, $\Lambda=1$ is an eigenvalue with an eigenvector in the direction of the flow. Its associated eigenspace is at least of dimension 2 since there is another eigenvector with eigenvalue 1 coming from the conserved quantity $H=E$. Therefore, according to the remark above, the orbit is elliptic if the spectrum of $J^T$ is $(1,1,{\mathrm e}^{i\omega},{\mathrm e}^{-i\omega})$ (and stable, except at some particular values), or hyperbolic if the spectrum is $(1,1,\lambda,1/\lambda)$ with $\lambda\in{\mathbb R}_*$ (unstable). The intermediate case is when the spectrum is restricted to $1$ or $-1$ and the orbit is called parabolic. Whether or not the parabolic periodic orbit is stable depends on higher order terms. In a more concise form, the above cases can be summarized using Greene's definition of a residue which led to a criterion on the existence of invariant tori~\cite{gree79,mack92}~:
$$
R=\frac{4-\mbox{tr}J^T}{4}.
$$
We notice that the 4 (instead of 2 for 2D maps) in the numerator comes from the two additional eigenvalues 1 coming from autonomous Hamiltonian flows.
If $R\in ]0,1[$, the periodic orbit is elliptic; if $R<0$ or $R>1$ it is hyperbolic; and if $R=0$ and $R=1$, it is parabolic and higher order expansions give the stability of such periodic orbits. Since the periodic orbit and its stability depend on the set of parameters ${\bm\alpha}$, the features of the dynamics will change with variations of the parameters. Generically, periodic orbits and their linear stability are robust to small changes of parameters, except at specific values where bifurcations occur. The proposed residue method to control chaos detects these rare events to yield the appropriate values of the parameters leading to the prescribed behavior on the dynamics. 

The residue method which leads to a reduction or an enhancement of the chaotic properties of the system is based on the change of stability of periodic orbits upon a change of the parameters of the system. For ${\bm\alpha}={\bf 0}$, let us consider two associated Birkhoff periodic orbits (i.e.\ periodic orbits having the same action but different angles in the integrable case and having the same rotation number on a selected Poincar\'e section), one elliptic ${\mathcal O}_e$ and one hyperbolic ${\mathcal O}_h$. Let us call $R_e$ and $R_h$ their residues. We have $R_e({\bf 0})>0$ (and smaller than one) and $R_h({\bf 0})<0$. We slightly modify the parameters ${\bm\alpha}$ until the elliptic periodic orbits becomes parabolic. Some particular situations arise at some critical value of the parameters ${\bm\alpha}={\bm\alpha}_c$~:

$(i)$~: $R_e({\bm \alpha}_c)=R_h({\bm \alpha}_c)=0$.\\

$(ii)$~: $R_e({\bm \alpha}_c)=0$ while $R_h({\bm \alpha}_c)<0$.\\

$(iii)$~: $R_e({\bm\alpha}_c)=1$ while $R_h({\bm \alpha}_c)<0$.\\

The first case is associated with the creation of an invariant torus. The two latter cases might be associated with the destruction of invariant tori (the ones around the elliptic periodic orbit). The third one is associated with a period doubling bifurcation. In this latter case, the change of stability of the new elliptic periodic orbit has to be considered.
Other interesting cases occur depending on the set of selected periodic orbits. 
The situation $(i)$ resembles the integrable situation where all the residues of periodic orbits of constant action are zero. It is expected that an invariant torus is reconstructed in this case. It can be associated with a transcritical bifurcation (an exchange of stability), a fold, or another type of bifurcation. In the situation $(ii)$, a change of stability occurs~: The elliptic periodic orbit turns hyperbolic while the hyperbolic one stays hyperbolic. It is generically characterized by a stationary bifurcation. In this case, the destruction of invariant curves is expected in general whether there are librational ones (representing the linear stability of an elliptic periodic orbit) or the neighboring rotational ones. 

An extra caution has to be formulated since this method only provides an indicator of the {\em linear} stability of periodic orbits. The nonlinear stability (or instability) has to be checked a posteriori by a Poincar\'e section for instance. This method only states that a bifurcation has occurred in the system, whether it is a stationary, transcritical, period doubling or other types of bifurcations. A more rigorous and safer control method would require to consider the global bifurcations, like the ones obtained by the intersections of the stable and unstable manifolds of two hyperbolic periodic orbits in the spirit of Ref.~\cite{olve87}. However such a control method would be computer-time consuming (determination of the stable and unstable manifolds) and hence not practical if some short time delay feedback is involved in the control process.  

\section{Application to a paradigmatic model}
\label{sec3}

\begin{figure}
\centering
\includegraphics[width=0.3\textwidth]{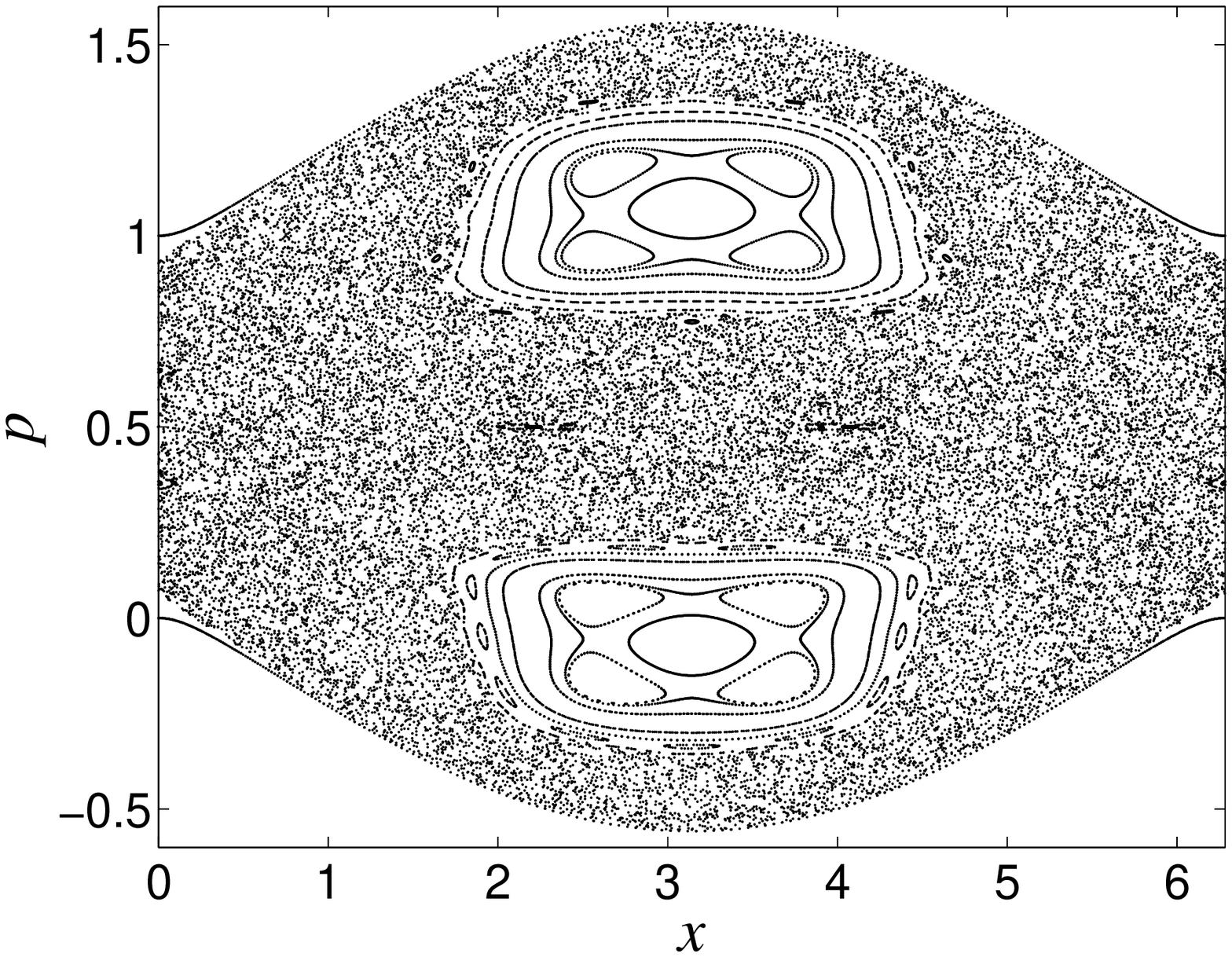}
\includegraphics[width=0.3\textwidth]{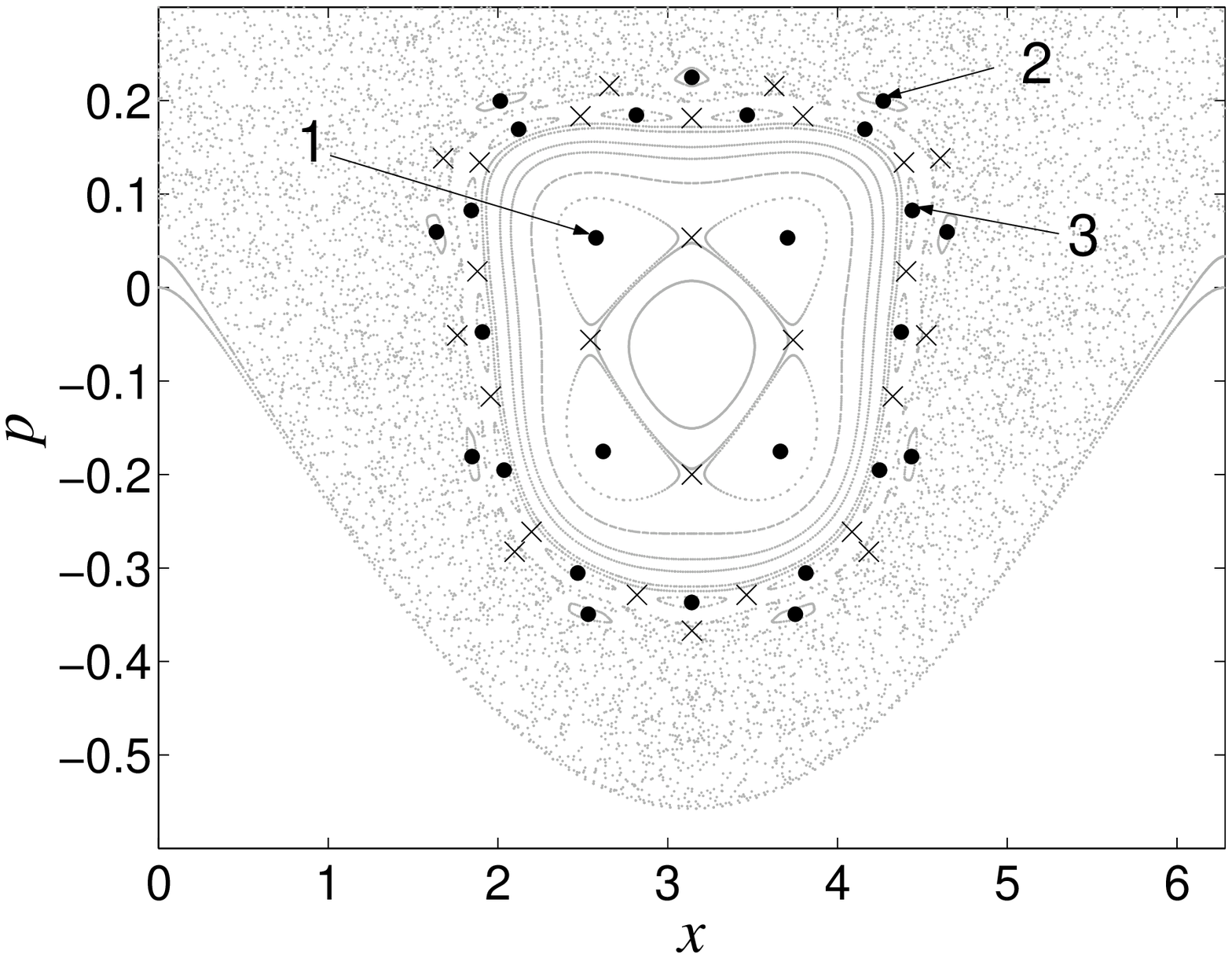}
\caption{Poincar\'e sections of Hamiltonian~(\ref{eqn:fp}) with $\varepsilon=0.065$. The arrows indicate the elliptic periodic orbits for the three cases considered here.}
\label{Fig1}
\end{figure}

\begin{figure}
\centering
\includegraphics[width=0.3\textwidth]{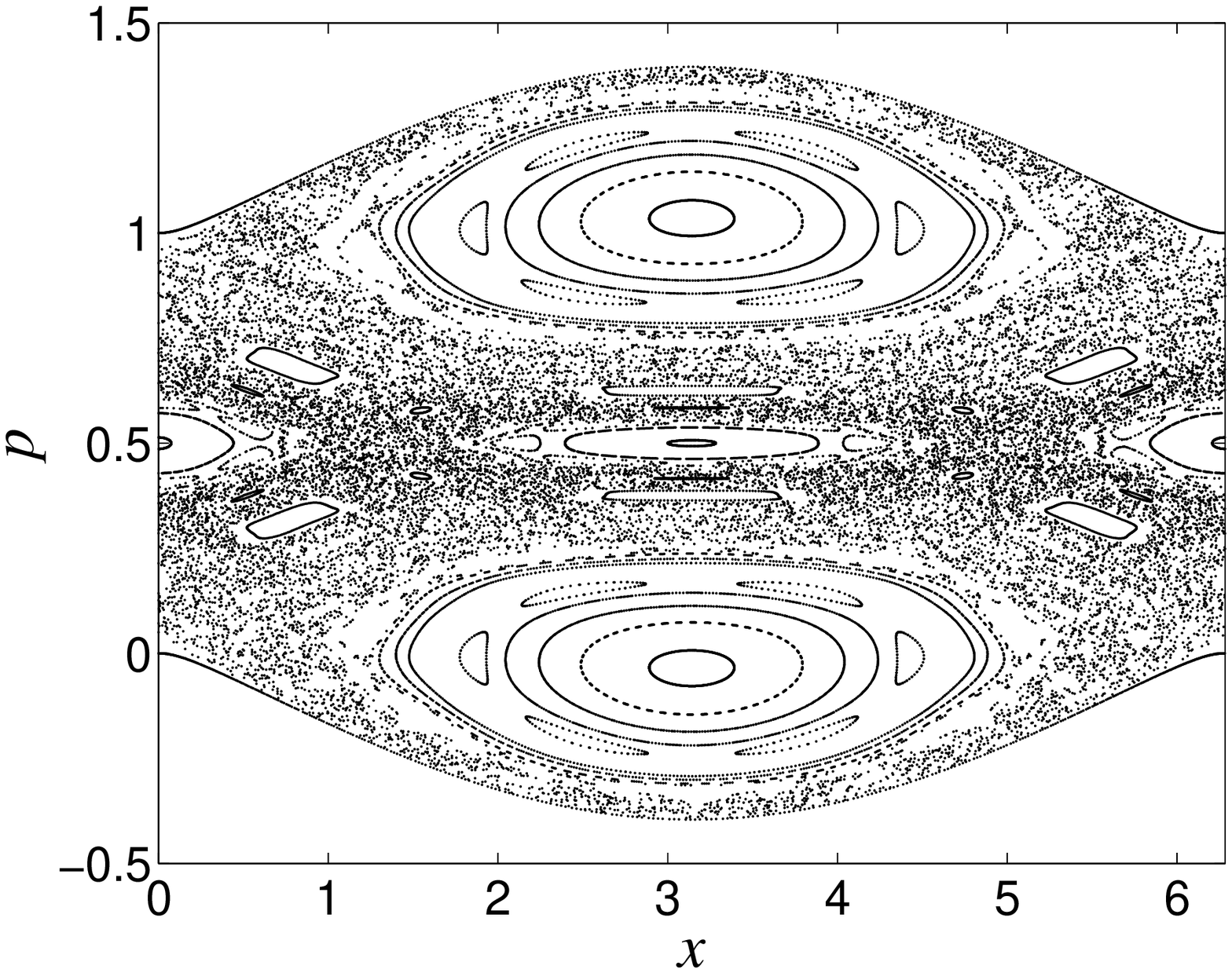}
\includegraphics[width=0.3\textwidth]{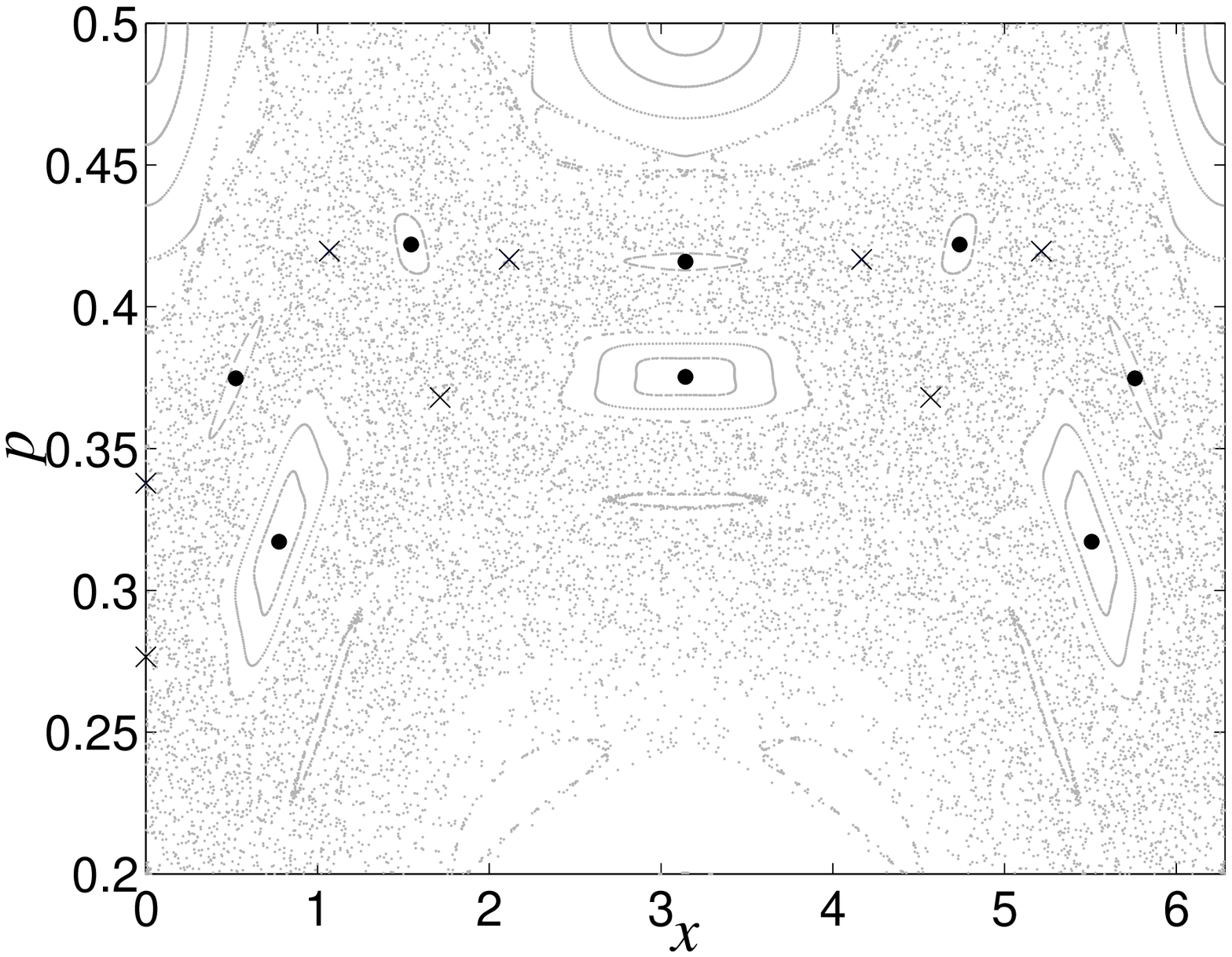}
\caption{Poincar\'e section of Hamiltonian~(\ref{eqn:fp}) for $\varepsilon=0.034$.}
\label{Fig2}
\end{figure}

We consider the following forced pendulum system with 1.5 degrees of freedom
\begin{equation}
\label{eqn:fp}
H(p,x,t)=\frac{p^2}{2}+\varepsilon\left( \cos x +\cos (x-t)\right).
\end{equation}
A Poincar\'e section of Hamiltonian~(\ref{eqn:fp}) is depicted on Fig.~\ref{Fig1} for $\varepsilon=0.065$ and on Fig.~\ref{Fig2} for $\varepsilon=0.034$.
In order to modify the dynamics of Hamiltonian~(\ref{eqn:fp}), we add an additional (control) parameter $k$~: 
We consider a family of Hamiltonians of the form
\begin{equation}
\label{eqn:fp2}
H_c(p,x,t)=\frac{p^2}{2}+\varepsilon\left( \cos x +\cos (x-t)\right)+\frac{k}{2}\varepsilon^2\cos(2x-t),
\end{equation}
where $k$ is not too large in order to consider a small modification of the original system, and minimizing the energy cost needed to modify the dynamics. Other choices of families of control terms are possible (not restricted to $k\cos (2x-t))$. In particular, more suitable choices of control terms would include more Fourier modes. We have selected a one-parameter family which originates from another control strategy which has been proved to be effective~\cite{chan05d}. The goal here is to determine the particular values of the parameter $k$ such that suitable modifications of the dynamics (which will be specified later) occur.

The algorithm is as follows~: First, we determine two periodic orbits of Hamiltonian~(\ref{eqn:fp}), an elliptic and a hyperbolic one with the same rotation number on the Poincar\'e section, using a multi-shooting Newton-Raphson method for flows~\cite{Bcvit05}. Then we modify continuously the control parameter $k$ and follow these two periodic orbits. We compute their residues as function of $k$. 

A first analysis is done on librational invariant tori (around the primary resonance located around $p\approx 0$). We point out in Fig.~\ref{Fig1} three particular elliptic periodic orbits (and their associated hyperbolic ones), labeled 1,2 and 3, with, respectively, $Q=4$, $Q=9$ and $Q=13$ intersections with the Poincar\'e section ($t=0 \mbox{ mod }2\pi$). These orbits will be used for two purposes~: First we follow the idea of Cary and Hanson on the construction of invariant tori. Then we extend the residue method to the destruction of these tori. A similar analysis is done on rotational invariant tori (the example of the goldenmean invariant torus is treated). This case allows us to compare the residue method with another approach on the control of Hamiltonian systems. 

Briefly we determine the control parameter $k$ such that there is a creation of an invariant torus if the original system does not have one, and the destruction of an invariant torus if the system does have one. 

\subsection{An exchange of stability associated with the creation of librational invariant tori}

\begin{figure}
\centering
\includegraphics[width=0.3\textwidth]{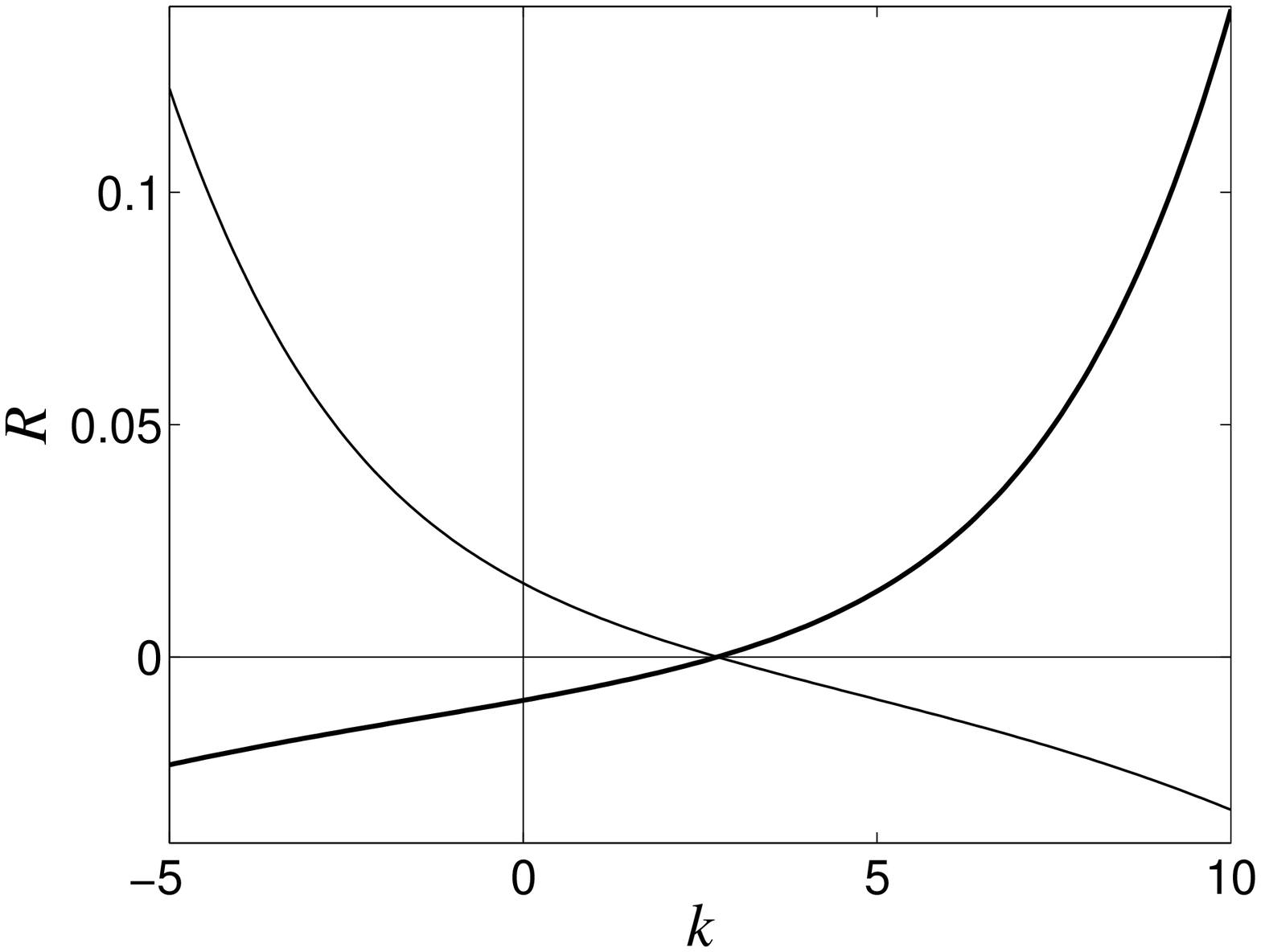}
\caption{Residues $R$ of the elliptic and hyperbolic (bold line) periodic orbits of Case 1 ($Q=4$) as functions of the parameter $k$ for Hamiltonian~(\ref{eqn:fp2}) with $\varepsilon=0.065$.}
\label{Fig3}
\end{figure}

For the case $Q=4$, Fig.~\ref{Fig3} represents the values of the residues of the elliptic and hyperbolic periodic orbits as functions of the control parameter $k$ [see Eq.~(\ref{eqn:fp2})]. At $k=k_c\approx 2.747$, both residues vanish which means that they become parabolic periodic orbits as in the integrable case. By increasing $k$, we notice that both orbits exchange their stability which is the manifestation of a transcritical bifurcation while each of the periodic orbits undergo individually a tangent bifurcation. This type of bifurcation has been observed in Refs.~\cite{blac90,erik93}. At $k=k_c$, an invariant torus is reconstructed. In order to check the robustness of the method, one could argue that since this invariant torus is composed of periodic orbits, it is not expected to be robust. However, by continuity in phase space, an infinite set of invariant tori is present in the neighborhood of the created invariant torus. Most of them have a frequency which satisfy a Diophantine condition and hence which will persist under suitable hypothesis on the type of perturbations. 

\begin{figure}
\centering
\includegraphics[width=0.3\textwidth]{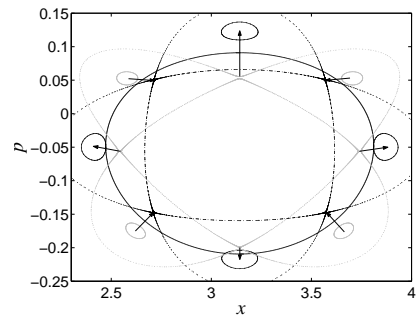}
\caption{Poincar\'e section of periodic orbits and some trajectories for Case 1 ($Q=4$) and Hamiltonian~(\ref{eqn:fp2}) with $\varepsilon=0.065$. The trajectories in gray are for $k=0$ and the ones in black are for $k=5$. At $k=k_c$, an invariant torus of the system is represented (bold line). The arrows indicate the change of locations of the periodic points as $k$ increases.}
\label{Fig4}
\end{figure}

The locations of the different periodic points on the Poincar\'e section as $k$ varies are indicated by arrows on Fig.~\ref{Fig4}. The change of stability of these periodic points is associated with the creation of an invariant torus (also represented in Fig.~\ref{Fig4} by the plot of the separatrices). We notice that apart from the exchange of stability, the phase space in the neighborhood of these periodic orbits is still regular (the chaotic region around the hyperbolic periodic orbits is not well developed), and hence the regular nature of phase space has not been changed locally (or one needs to consider higher values of the parameters).

\begin{figure}
\centering
\includegraphics[width=0.3\textwidth]{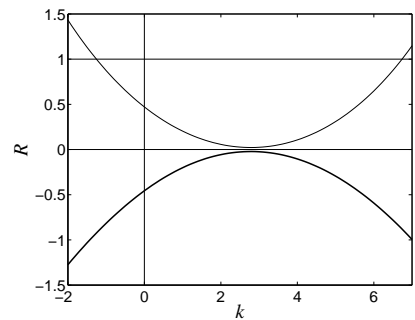}
\caption{Residues $R$ of the elliptic and hyperbolic periodic orbits of Case 2 ($Q=9$) as functions of the parameter $k$ for Hamiltonian~(\ref{eqn:fp2}) with $\varepsilon=0.065$.}
\label{Fig5}
\end{figure}

The same analysis can be carried out on a set of periodic orbits which are located in a more chaotic region, like for instance the two cases $Q=9$ and $Q=13$ of Fig.~\ref{Fig1} outside the regular resonant island. The values of the residues as functions of the parameter $k$ are respectively represented on Figs.~\ref{Fig5} and \ref{Fig6} for $Q=9$ and $Q=13$. For $Q=9$, we notice that the residues do not vanish in the range of $k$ we considered although there are small and extremum at the same value of the parameter $k\approx 2.79$. However, even if these residues do not vanish (and therefore no exchange of stability by the creation of an invariant torus), there is a significant regularization of the dynamics at this specific value of the parameter (not shown here).
For $Q=13$, the residues vanish for $k\approx 2.76$ (see Fig.~\ref{Fig6}) and there is a transcritical bifurcation associated with the creation of a set of invariant tori like in Figs.~\ref{Fig3} and \ref{Fig4}. The associated phase space shows a significant increase of the size of the resonant island

\subsection{Enhancing chaos near a resonant island}

\begin{figure}
\centering
\includegraphics[width=0.3\textwidth]{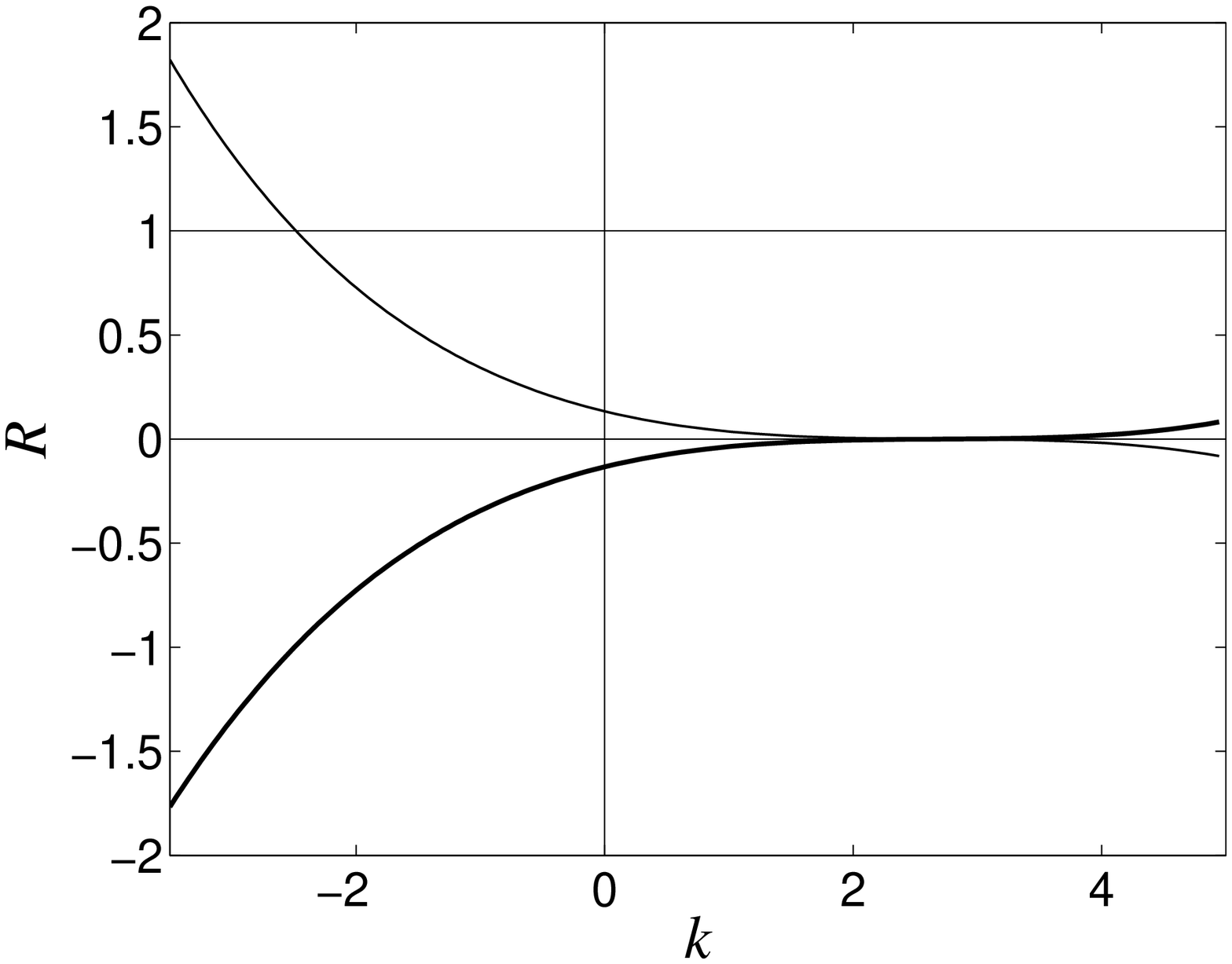}
\caption{Residues $R$ of the elliptic and hyperbolic periodic orbits of Case 3 ($Q=13$) as functions of the parameter $k$ for Hamiltonian~(\ref{eqn:fp2}) with $\varepsilon=0.065$.}
\label{Fig6}
\end{figure}

\begin{figure}
\centering
\includegraphics[width=0.3\textwidth]{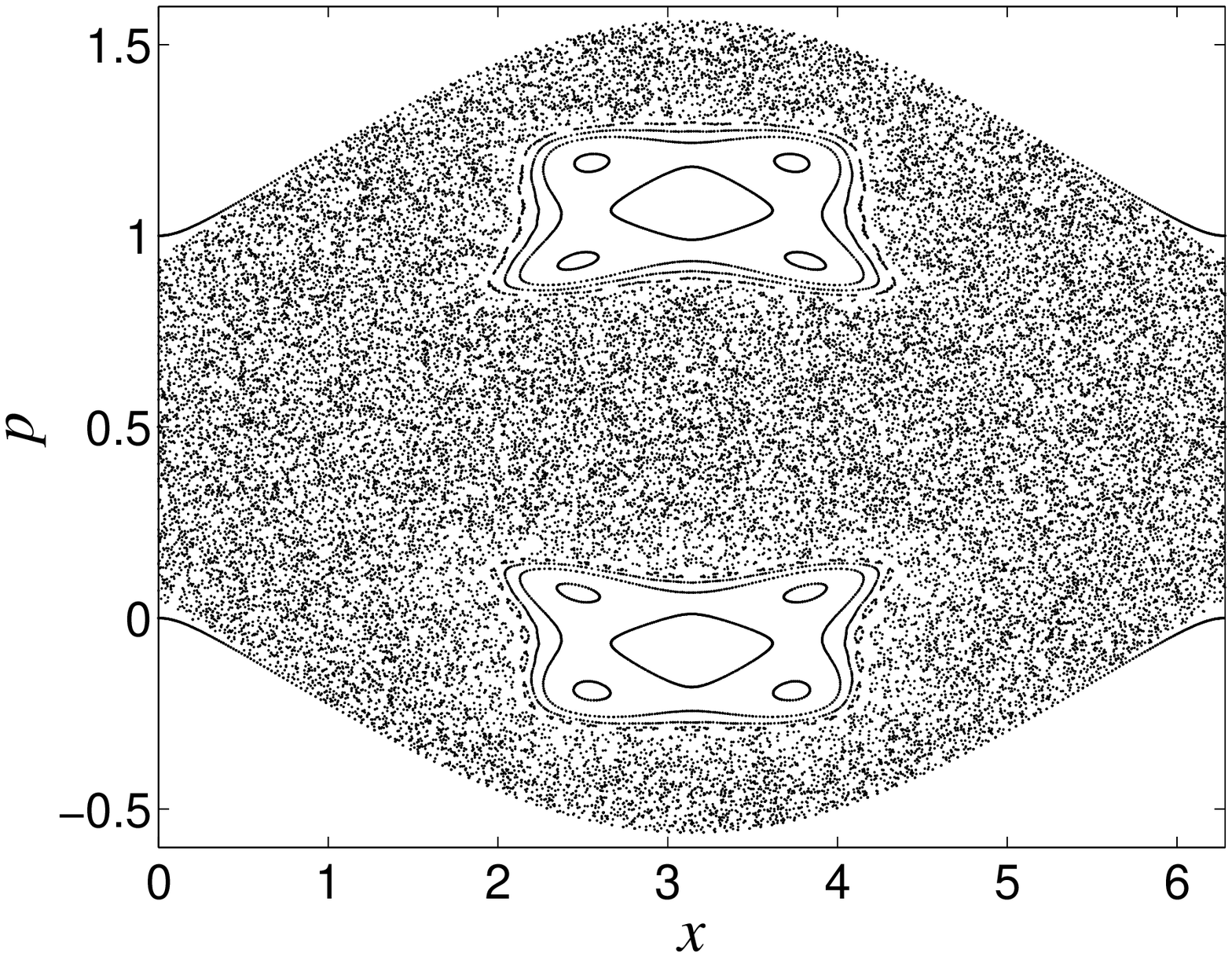}
\includegraphics[width=0.3\textwidth]{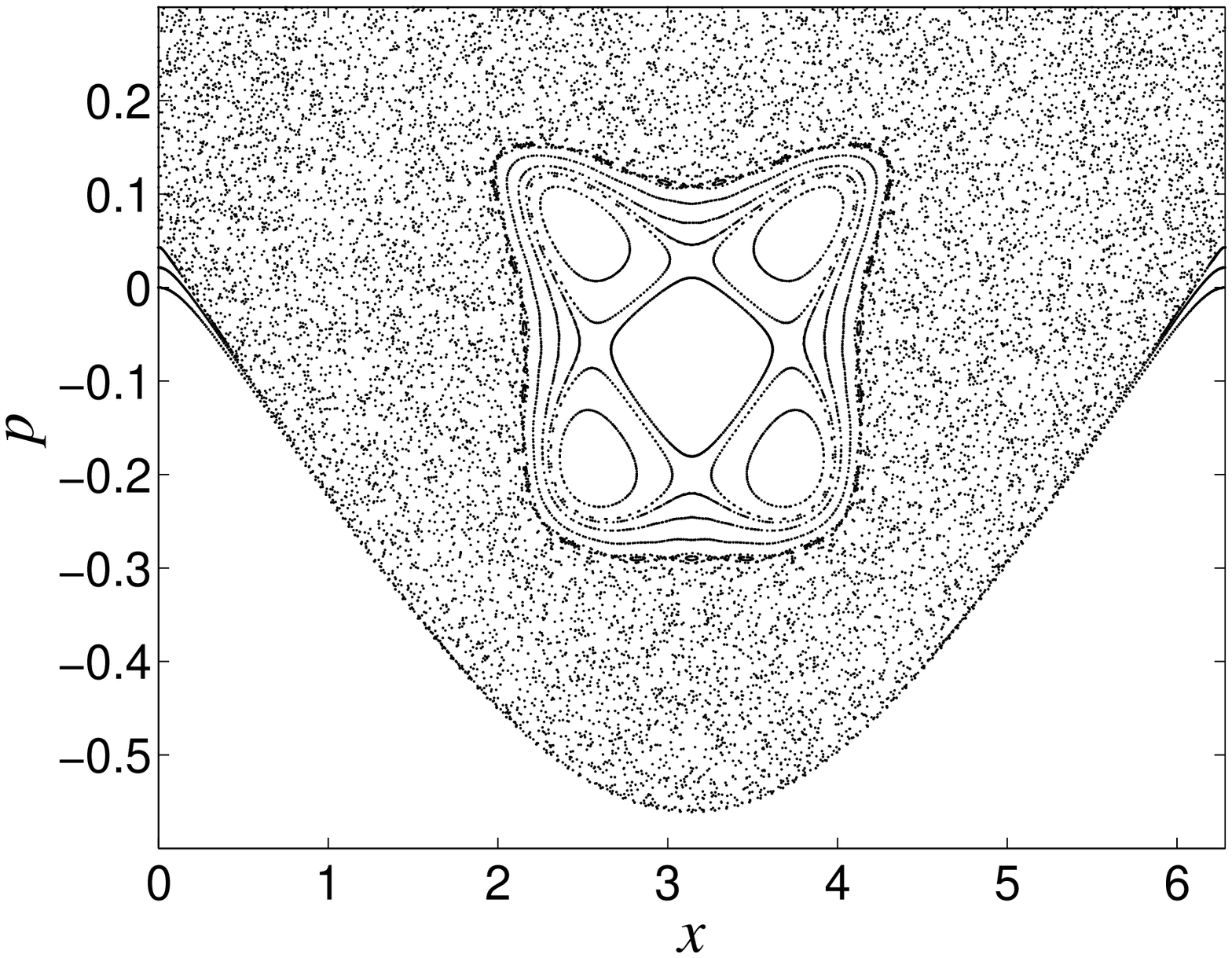}
\caption{Poincar\'e section of Hamiltonian~(\ref{eqn:fp2}) for $\varepsilon=0.065$ and $k=-2.484$.}
\label{Fig7}
\end{figure}

In this section, we address the destruction of a resonant island by breaking up librational invariant tori.
We notice that on Fig.~\ref{Fig5}, a bifurcation occurs at $k\approx -1.254$ for the Case 2 ($Q=9$) when the residue of the elliptic periodic orbit becomes equal to 1. The neighborhood of this periodic orbit becomes a chaotic layer and the qualitative change in the dynamics is seen since the chaotic layer becomes thicker at this value of the parameter. However since this periodic orbit was initially (at $k=0$) already in the outer chaotic region (see Fig.~\ref{Fig1}),  the regularization is not drastic. 
In order to obtain a more significant change in the dynamics and a large chaotic zone, one needs to select a periodic orbit inside a regular region, like for instance the one with $Q=13$. 
A bifurcation occurs at $k\approx -2.484$ where the residue of the elliptic periodic orbit crosses 1 (see Fig.~\ref{Fig6}). A Poincar\'e section for the latter case is depicted on Fig.~\ref{Fig7}, and shows that a significantly large neighborhood has been destabilized by the control term. We notice that the ratio between the size of the control term and the one of the perturbation is equal to $k\varepsilon\approx 0.16$. We also notice that this last value of $k$ is larger than the one required for $Q=9$. As expected, one needs a larger amplitude to destabilize a region closer to a regular one. A more effective destabilization procedure can be obtained with the periodic orbit $Q=4$ which is inside the regular region. However, as mentioned, the value necessary for this destabilization ($k\approx -12.5$ or for $k\approx 16.3$) is too large; hence we discard it because of our restriction on energy cost.

\subsection{Creation of the goldenmean rotational invariant torus}

In this section, we apply the same approach on rotational invariant tori. It allows us to compare the results with the ones obtained by a control method proposed in Refs.~\cite{vitt05,chan05b,chan05d}. First, the idea is to look at the creation of a specific invariant torus. For instance, we select a torus which has been widely discussed in the literature (see Ref.~\cite{chan02} and references therein), the goldenmean one, which has a frequency $\omega=(3-\sqrt{5})/2$ for Hamiltonian~(\ref{eqn:fp2}). We choose $\varepsilon=0.034$, and we first notice that when $k=0$, this Hamiltonian does not have such an invariant torus (since its critical value is $\varepsilon\approx 0.02759$~\cite{chan02}). The purpose here is to find the value of the control parameter $k$ needed by the residue method to reconstruct this invariant torus (such that Hamiltonian~(\ref{eqn:fp2}) has this invariant torus).

The idea of doing this follows Greene's residue criterion. By performing an appropriate change of stability on higher and higher order periodic orbits, the amplitude of the control term should be smaller and smaller.
For $Q=2$, the residues of the elliptic and hyperbolic periodic orbit vanish at $k\approx 4.3$, and for $Q=3$, at $k\approx 3.4$.   
For $Q=5$, both residues vanish at $k=2.98$ and also at $k\approx 5.07$. At these values of the parameter, the phase space is locally filled by invariant tori where it is also expected that the goldenmean invariant torus is present. We notice that the elliptic periodic orbit with $Q=8$ (the next one in Greene's residue approach for the analysis of the golden mean torus) is destabilized at $\varepsilon\approx 0.0325$. Therefore, there is no elliptic periodic orbit with $Q=8$ at $\varepsilon=0.034$ and the analysis using the coupled elliptic/hyperbolic periodic orbits cannot be carried out. However, by following the two (initially hyperbolic) periodic orbits with $Q=8$, we see that both residues vanish at $k\approx 2.84$. 

We compared these values of stabilization with the one given by a method of local control based on an appropriate modification of the potential to reconstruct a specific invariant torus~\cite{vitt05,chan05d}. Such method provides explicitly the shape (and amplitude) of possible control terms whereas the one used in this article has been guessed from these references. By appropriate truncation (keeping the main Fourier mode), this method provides 
$$f(x,t)=\frac{\varepsilon^2}{2\omega(1-\omega)}\cos (2x-t),$$ 
where $\omega=(3-\sqrt{5})/2$, as an approximate control term. Therefore the amplitude is $k=1/\omega(1-\omega) \approx 4.24$ which is of the same order as the values obtained by zeroing the residues. However, we point out that smaller values are obtained by looking at higher periodic orbits. 

Therefore, an efficient control strategy is to combine the advantages of both methods~: First, the specific shape of the terms that have to be added to regularize the system is obtained using the method of Ref.~\cite{chan05d}. Then the amplitudes of these terms are lowered using high order periodic orbits. By considering the control term used in this article, we expect that zeroing the residues of high period will not be feasible with just this term (as it is the case for instance in Fig.~\ref{Fig5}). A more suitable form of control terms would be constructed from an exact control term which is
\begin{eqnarray*}
f(x,t)&=&\frac{\varepsilon^2}{2\omega(1-\omega)}\cos (2x-t)\\
&&-\frac{\varepsilon^2}{4\omega^2}\cos 2x-\frac{\varepsilon^2}{4(1-\omega)^2}\cos 2(x-t).
\end{eqnarray*}
However, it should be noticed that a control term given by Ref.~\cite{chan05d} is not always experimentally accessible. The idea is to use a projection of this control term onto a basis of accessible functions. This projected control term would give an idea of the type of control terms to be used for the residue method. 

We would like to stress that in the absence of elliptic islands an initial guess for the Newton-Raphson method is not straightforward from the inspection of the Poincar\'e section. In particular, it is not easy to select the appropriate hyperbolic periodic orbits which will lead to a significant change in the dynamics. However, once it has been located, the method can follow them by continuity in the same way as the elliptic ones since the Newton-Raphson method does not depend on the linear stability of these orbits. This makes the method more difficult (although possible) to handle for just hyperbolic periodic orbits.

\subsection{Destruction of the goldenmean rotational invariant torus}

\begin{figure}
\centering
\includegraphics[width=0.3\textwidth]{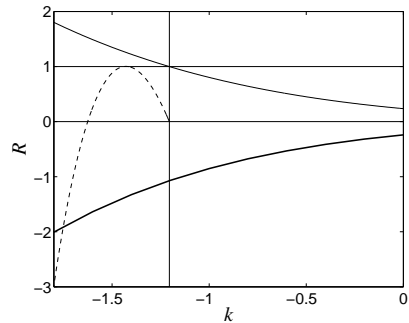}
\caption{Residues $R$ of the elliptic and hyperbolic periodic orbits with $Q=13$ and also of the one with $Q=26$ (dashed line) born out of a period doubling bifurcation for Hamiltonian~(\ref{eqn:fp2}) with $\varepsilon=0.0275$.}
\label{Fig8}
\end{figure}

In this section, we consider Hamiltonian~(\ref{eqn:fp2}) with $\varepsilon=0.0275$. We notice that for $k=0$, Hamiltonian~(\ref{eqn:fp2}) does have the rotational goldenmean torus. The purpose is to find some small values of the parameter $k$ where this invariant torus is destroyed. We notice that this case is easier to find than in the previous section since it is well-known that any additional perturbation will end up by destroying an invariant torus generically. Here it means that there will be large intervals of parameters for which the torus is broken (contrary to the case of the creation of invariant tori). However we will add an additional assumption that the parameters for which this invariant torus is destroyed has to be small compared with the perturbation. 
We also notice that the destruction of the golden mean invariant torus is first obtained for negative values of the control parameter (see Fig.~\ref{Fig8}).

First we illustrate the method by considering specific elliptic and hyperbolic periodic orbits (with winding ratio $5/13$) near the goldenmean torus which will show the changes of dynamics occurring as the parameter is varied. We notice that the behaviors described below are generic for all the neighboring periodic orbits.

\begin{figure}
\begin{center}
\unitlength 1cm
\begin{picture}(10,5)
\put(0,0){\includegraphics[width=0.3\textwidth]{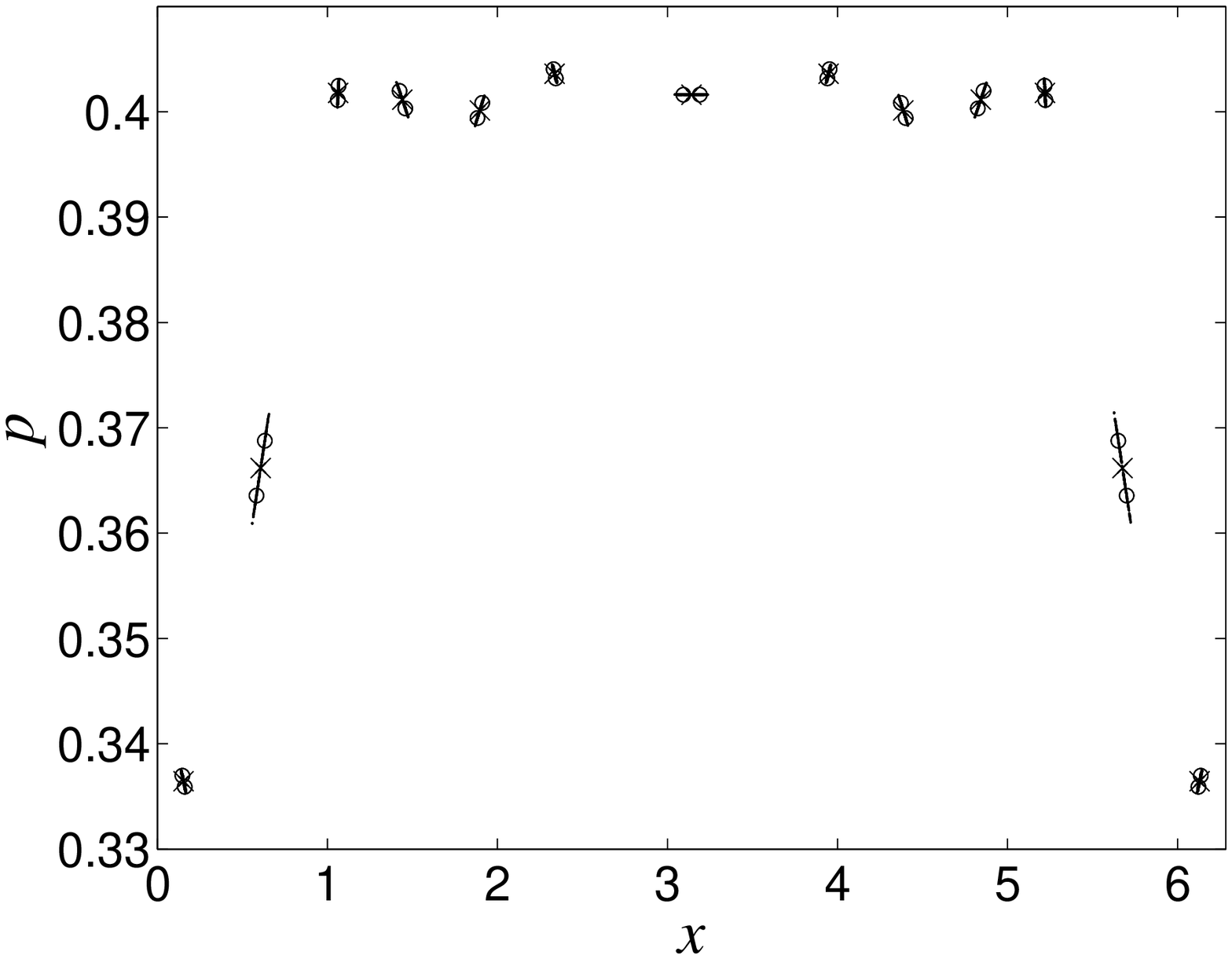}}
\put(1.5,0.7){\includegraphics[width=0.17\textwidth]{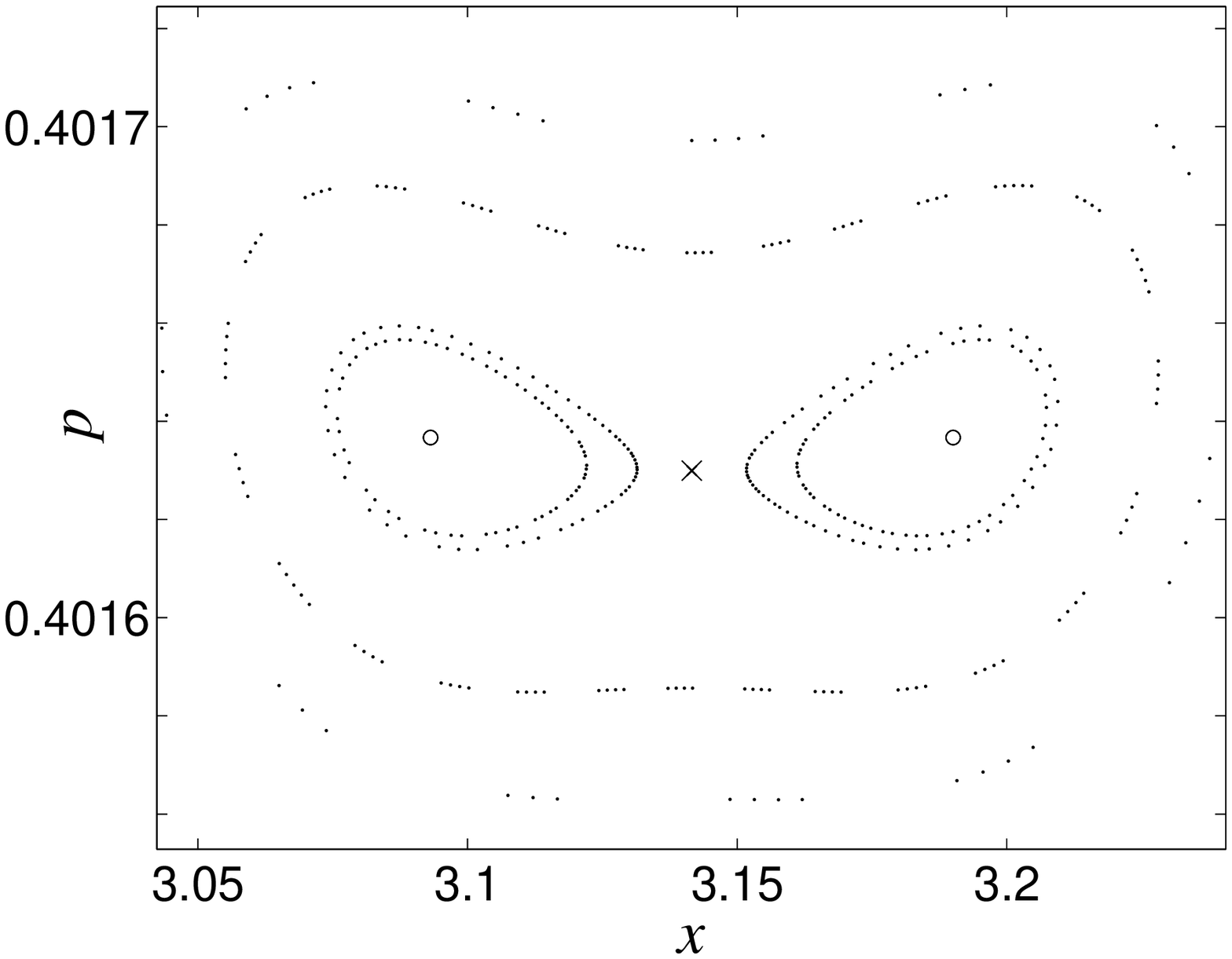}}
\end{picture}
\end{center}
\caption{Poincar\'e section around the periodic orbit with winding ratio $5/13$ (indicated with crosses) for Hamiltonian~(\ref{eqn:fp2}) for $\varepsilon=0.0275$ and $k=-1.215$. The period orbit with period 26 (indicated by circles) results from a period doubling bifurcation of the one with period 13 (represented by crosses on the Poincar\'e section).}
\label{Fig9}
\end{figure}

The residues of these periodic orbits as functions of the parameter $k$ are shown in Fig.~\ref{Fig8}. We notice that
the elliptic periodic orbit changes its stability, i.e.\ becomes hyperbolic, at $k\approx -1.205$ (where its residue becomes equal to 1). A close inspection of the Poincar\'e section shows on Fig.~\ref{Fig9} that it undergoes a period doubling bifurcation into an elliptic periodic orbit with 26 intersections on the Poincar\'e section (and winding ratio $10/26$) which has a residue zero at the bifurcation. By following the residue of this elliptic periodic orbit (depicted by a dashed line in Fig.~\ref{Fig8}) we see that it vanishes for $k=-1.6256$. At this value of the parameter and for higher value in amplitude, all the periodic orbits considered here (the two with $Q=13$ and the one with $Q=26$) are hyperbolic. Therefore it is expected that there is a chaotic zone in this area and it is a value at which the torus is expected to be broken (confirmed by a close inspection of the Poincar\'e section).

It is important to notice that a vanishing residue does not automatically imply that there is a creation of an invariant torus, contrary to the previous cases which were obtained by using jointly the elliptic and hyperbolic periodic orbits (and vanishing residues in both cases). Here the hyperbolic periodic orbit associated with these elliptic periodic orbits (which is the periodic orbit from which the new elliptic orbit was born out by a period doubling bifurcation) stays hyperbolic as the residue of the elliptic one vanishes. This feature is generic~:
The same analysis has been carried out for
 higher order elliptic periodic orbits close to the goldenmean invariant torus, i.e.\ the ones with winding ratio $8/21$, $13/34$, $21/55$, $34/89$~: First, the values of the control parameter for which the residues (which are around $0.25$ for $k=0$ and increase as $k$ decreases) cross 1 are computed and reported in Table~I (denoted $k(R=1)$). At these values of the parameters, a period doubling bifurcation occurs for each of them. Then we follow the residues of the elliptic periodic orbits with double period $Q=42$, $Q=68$, $Q=110$ and $Q=178$. The parameter values at which these residues vanish are also reported in Table~I. For instance, using the periodic orbit with winding ratio $8/21$, we obtain $k=-0.935$ as the value at which the residue of the bifurcated elliptic periodic orbit with winding ratio $16/42$. If we consider higher order periodic orbits, it happens that the goldenmean invariant torus is destroyed by this additional perturbation but not the ones in the neighborhood. If one is looking at large scale transport properties, these other invariant tori have to be taken into account.

\begin{table}

	\centering
		\begin{tabular}{|c||c|c|c|c|c|}
		\hline
			$Q$ & 13 & 21 & 34 & 55 & 89\\\hline
			$k(R=1)$ & -1.205 & -0.705 & -0.435 & -0.273 & -0.179\\\hline
			$k_c$ & -1.626 & -0.935 & -0.566 & -0.350 & -0.225 \\\hline
		\end{tabular}
		
		\caption{Values of the parameter $k$ at which the residue of the elliptic periodic orbit with period $Q$ crosses 1 (denoted $k(R=1)$) and at which the residue of the elliptic periodic orbit with period $2Q$ obtained by period doubling bifurcation at $k(R=1)$ vanishes (denoted $k_c$).  }
\end{table}

\section*{Concluding remarks}
In this article, we reviewed and extended a method of control of Hamiltonian systems based on linear stability analysis of periodic orbits. We have shown that by varying the parameters such that the residues of selected periodic orbits cross 0 or 1, some important bifurcations happen in the system. These bifurcations can lead to the creation or the destruction of invariant tori, depending on the situation at hand. Therefore we have proposed a possible extension of the residue method to the case of increasing chaos locally. Moreover, we have compared two methods of chaos reduction, and by taking advantage of both methods, we have devised a more effective control strategy.
It is worth noticing that the extension of Cary-Hanson's method to four dimensional symplectic maps has been done in Refs.~\cite{wan98,wan01} for the increase of dynamic aperture in accelerator lattices. The extension to the destruction of invariant surface would be to consider the change of linear stability of selected periodic orbits. However, it would require to consider new types of bifurcations which occurs in the system, like for instance, Krein collisions~\cite{howa87}.

\begin{acknowledgments}
This work is supported by Euratom/CEA (contract EUR 344-88-1 FUA F).
We acknowledge useful discussions, comments and remarks from J.R. Cary and the Nonlinear Dynamics group at CPT. 
\end{acknowledgments}

%\bibliographystyle{unsrt}
%\bibliography{BiblioTeX}

\begin{thebibliography}{10}

\bibitem{Bcvit05}
P.~Cvitanovi\'c, R.~Artuso, R.~Mainieri, G.~Tanner and G.~Vattay,
{\em Chaos: {C}lassical and {Q}uantum}
(Niels Bohr Institute, Copenhagen, 2005),
archived in \texttt{http://ChaosBook.org}.

\bibitem{hans84}
J.D. Hanson and J.R. Cary, {\em Elimination of stochasticity in stellarators},
Phys. Fluids {\bf 27}, 767 (1984).

\bibitem{cary86}
J.R. Cary and J.D. Hanson,
{\em Stochasticity reduction},
Phys. Fluids {\bf 29}, 2464 (1986).

\bibitem{gree79}
J.M. Greene, {\em A method for determining a stochastic transition},
J. Math. Phys. {\bf 20}, 1183 (1979).

\bibitem{hans94}
J.D. Hanson, {\em Correcting small magnetic field non-axisymmetries},
Nuclear Fusion {\bf 34}, 441 (1994).

\bibitem{wan98}
W.~Wan and J.R. Cary, {\em Increasing the dynamic aperture of accelerator lattices},
Phys. Rev. Lett. {\bf 81}, 3655 (1998).

\bibitem{mack92}
R.S. MacKay, {\em Greene's residue criterion},
Nonlinearity {\bf 5}, 161 (1992).

\bibitem{olve87}
A.~Olvera and C.~Sim\'o, {\em An obstruction method for the destruction of invariant curves},
Physica {\bf 26D}, 181 (1987).

\bibitem{chan05d}
C.~Chandre, M.~Vittot, G.~Ciraolo, Ph. Ghendrih and R. Lima,
{\em Control of stochasticity in magnetic field lines},
Nuclear Fusion {\bf 46}, 33 (2005).

\bibitem{blac90}
R.C. Black and I.I. Satija,
{\em Universal pattern underlying the recurrence of
  Kolmogorov-Arnold-Moser tori},
Phys. Rev. Lett. {\bf 65}, 1 (1990).

\bibitem{erik93}
A.B. Eriksson and P.~Dahlqvist,
{\em Stability exchanges between periodic orbits in a Hamiltonian
  dynamical system},
Phys. Rev. E {\bf 47}, 1002 (1993).

\bibitem{vitt05}
M.~Vittot, C.~Chandre, G.~Ciraolo and R.~Lima.
{\em Localized control for non-resonant {H}amiltonian systems},
Nonlinearity {\bf 18}, 423 (2005).

\bibitem{chan05b}
C.~Chandre, G.~Ciraolo, F.~Doveil, R.~Lima, A.~Macor and M.~Vittot,
{\em Channeling chaos by building barriers},
Phys. Rev. Lett. {\bf 94}, 074101 (2005).

\bibitem{chan02}
C.~Chandre and H.R. Jauslin,
{\em Renormalization-group analysis for the transition to chaos in
  {H}amiltonian systems},
Phys. Rep. {\bf 365}, 1 (2002).

\bibitem{wan01}
W.~Wan and J.R. Cary, {\em Method for enlarging the dynamic aperture of accelerator lattices},
Phys. Rev. Spec. Top.--Accel. Beams {\bf 4}, 084001 (2001).

\bibitem{howa87}
J.E. Howard and R.S. MacKay, {\em Linear stability of symplectic maps},
J. Math. Phys. {\bf 28}, 1036 (1987).

\end{thebibliography}

\end{document}